\documentstyle[epsf,12pt]{article}
\newcommand{\Gl}[2]{{\mbox{Gl}}_{#1}\left(#2\right)}
\newcommand{\Li}[2]{{\mbox{Li}}_{#1}\left(#2\right)}
\newcommand{\Cl}[2]{{\mbox{Cl}}_{#1}\left(#2\right)}
\newcommand{\Ls}[2]{{\mbox{Ls}}_{#1}\left(#2\right)}
\newcommand{\LS}[3]{{\mbox{Ls}}_{#1}^{(#2)}\left(#3\right)}

\begin{document}
\begin{center}

\begin{flushright}
BI-TP 99/34
\end{flushright}

\vspace{15mm}

{\large \bf  Single mass scale diagrams: construction of
a basis for the $\varepsilon$-expansion.}

\vspace{15mm}

{\large
J.~Fleischer$^a$
\footnote{E-mail: fleischer@physik.uni-bielefeld.de},~
M.~Yu.~Kalmykov$^{a,b}$
\footnote{~E-mail: misha@physik.uni-bielefeld.de}
}

\vspace{15mm}

\begin{itemize}
\item[$^a$]
{\it ~Fakult\"at f\"ur Physik, Universit\"at Bielefeld,
      D-33615 Bielefeld, Germany}
\item[$^b$]
{\it BLTP, JINR, 141980, Dubna (Moscow Region), Russia}
\end{itemize}
\end{center}

\begin{abstract}
Exploring the idea of Broadhurst on the ``sixth root of unity'' we present 
an ansatz for
construction of a basis of transcendental numbers for the 
$\varepsilon$-expansion of single mass scale diagrams with two particle
massive cut. As example, several new two- and three-loop master integrals 
are calculated.
\end{abstract}


\vspace*{10pt}

{\it PACS number(s)}: 12.38.Bx

\thispagestyle{empty}
\setcounter{page}0
\newpage

\section{Introduction}

One of the interesting cases of diagrams occurring in QCD and the 
Standard Model is of the propagator type with one mass and the external
momentum on the mass shell (the bubble (vacuum) type diagrams also belong to this
class). Only a limited number of analytical results is available in this
case: the two- and three-loop diagrams occurring in QCD are given in
Refs.\cite{broadhurst1,broadhurst2}; the analytical result for the two-loop 
bubble integral are presented up to the finite part in 
\cite{broadhurst1,tadpole,master} and the $O(\varepsilon)$ term in \cite{magic};
the finite part of three-loop bubble master integrals is calculated in 
\cite{master3}; the results
for all other two-loop on-shell propagator type integrals with one mass
are collected in \cite{master2}. For some diagrams the results are available
in terms of hypergeometric functions \cite{one-loop,two-loop}. However,
the $\varepsilon$-expansion of hypergeometric functions is an independent
and sometimes very complicated task. 

At the same time a lot of different methods for numerical calculations
of Feynman diagrams are available now
\cite{fv3,ghinculov1,Kreimer,small,Pivovarov}. From another point of 
view the development of numerical algorithms, in
particular PSLQ \cite{pslq}, allows to determine with high confidence,
whether or not particular numerical results for master-integrals can be
expressed as a rational linear combination of several given transcendental
constants. However, the problem arises of how to define the full set of
the transcendental structure (the ``basis'') occurring in given
diagrams. The structure of massless multiloop integrals is well
understood at the present time. In contains mainly the $\zeta$-function
\cite{massless1} and Euler sums \cite{massless2}. Recently the
connection between knot theory and the transcendental structure of Feynman
diagrams was established \cite{knot}. In more complicated case, when
dimensionless parameters exist, harmonic sums \cite{FKV,harmonic} 
appeared. For studying the massive case, the differential equation
method \cite{kotikov} and geometrical approach 
\cite{geometrical,davydychev1} can be useful. 

The aim of our investigation is to find the transcendental structure of
single mass scale propagator type diagrams. In this paper we
restrict ourselves to the consideration of diagrams with two particle
massive cuts only. All our results are obtained empirically by
carefully compiling and examining a huge data base on high precision
(about 200 decimals) numerical calculation \cite{pslq}.

\section{Basis}

\begin{figure}[th]
\centerline{\vbox{\epsfysize=30mm \epsfbox{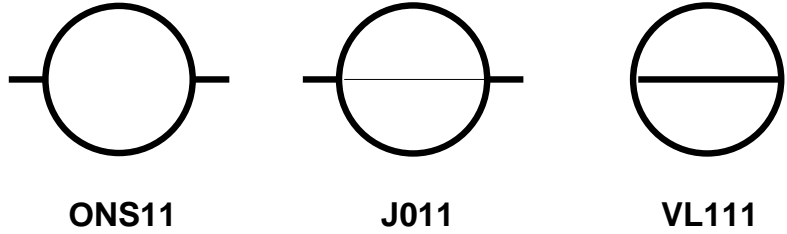}}}
\caption{\label{simple} Examples of the simplest single mass scale
integrals with two-particle massive cut.
Bold and thin lines correspond to massive and
massless propagators, respectively.}
\end{figure}

Our aim is to construct a basis (the full set of transcendental numbers)
for single mass scale propagator type diagrams with two-particle
massive cuts. The simplest diagrams of this type are shown in
Fig.\ref{simple}.

An algebra for such a basis can be constructed as follows: the product of 
two basis elements
having weights $J_1$ and $J_2$, respectively, must belong to the basis of
weight $J_1+J_2$. The weight of basis elements can be connected, e.g., 
with powers of $\varepsilon$ within the Laurent expansion of dimensionally
regularized \cite{dim} Feynman diagrams. 

In constructing the basis of weight J we
will use Broadhurst's observation \cite{master3}, that the sixth root of
unity plays an important role in the calculation of the diagrams (see also
\cite{trans}). Accordingly we conjecture, that the basis of weight J contains 
functions obtained by splitting of polylogarithms  
$\Li{J}{z}, \Li{J}{1-z}$ into real and imaginary parts, where 
$z=exp\left( i \frac{\pi}{3}\right),exp\left( i \frac{2 \pi}{3}\right).$
The general case, $z=exp\left( \pm i \frac{\pi}{3}+ i \pi m \right)$ 
can be reduced to these two values \cite{Lewin}. E.g., 
with $\Li{2n}{e^{i \theta}} = \Gl{2n}{\theta} + i \Cl{2n}{\theta}$, 
the basis of weight $2n$ must contain $\Gl{2n}{\frac{\pi}{3}},
\Gl{2n}{\frac{2\pi}{3}}, \Cl{2n}{\frac{\pi}{3}}$ and
$\Cl{2n}{\frac{2\pi}{3}}$. Not all of these functions are independent:
$\Gl{2n}{\omega \pi} \sim \zeta_{2n}$, where $\omega$ is an arbitrary
rational number; $\left(1+2^{1-2n} \right) \Cl{2n}{\frac{2\pi}{3}} = 
\Cl{2n}{\frac{\pi}{3}}$. 

The basis we are going to construct is supposed to hold for any number
of loops and arbitrary order of the $\varepsilon$-expansion. 
Consequently we connect the lowest weight basis elements with
$\Li{1}{z}$ and $\Li{1}{1-z}$, which results in $\pi$ and $\ln 3$. 
Comparing this with the one-loop diagram of Fig.\ref{simple} ({\bf ONS11}),
the expression given in \cite{davydychev1} allows us to find from its finite 
part the transcendental normalization factor $\frac{1}{\sqrt{3}}$, so 
that our lowest basis is given by $\frac{\pi}{\sqrt{3}}$ and $\ln 3$. 
In general the normalization factor $\frac{1}{\sqrt{3}}$ is introduced
according to our experience.

Using the above conjectures, we obtain the basis of higher weights: for
weight {\bf 2}, e.g., we have 
$\frac{\pi}{\sqrt{3}} \ln 3, \zeta_2$, $\frac{1}{\sqrt{3}}\Cl{2}{\frac{\pi}{3}}$
and $\ln^2 3$. The results up to weight ${\bf 5}$ are presented in Table I.
Some transcendental structures like, e.g., $\Ls{3}{\frac{\pi}{3}}$ 
do not appear in the table since they could be 
eliminated by means of relations given in \cite{Lewin}. Including weight {\bf 3}
the relations found in \cite{Lewin} were sufficient. For weight {\bf 4} we needed
the new relation

\begin{equation}
\Cl{4}{\frac{\pi}{3}}  =  \frac{2}{9} \Ls{4}{\frac{\pi}{3}} 
- \frac{1}{9} \pi \zeta(3),
\end{equation}

\noindent
and for weight {\bf 5} we used the further relations

\begin{eqnarray}
\LS{5}{1}{\pm \frac{\pi}{3}} & = & \frac{1}{3} \pi \Ls{4}{\frac{\pi}{3}}
-\frac{19}{4} \zeta_5 - \frac{1}{2} \zeta_2 \zeta_3,
\\
\LS{5}{2}{\pm \frac{\pi}{3}} & = & \pm \left[
\frac{2}{3} \Ls{5}{\frac{\pi}{3}}
+\frac{253}{54} \pi \zeta_4 \right],
\end{eqnarray}

\noindent
in order to restrict the basis as much as possible.

$$
\begin{array}{|c|c|c|c|c|||c|} \hline
\multicolumn{6}{|c|}{Table ~~~I }    \\   \hline
{\bf 1}  & {\bf 2}  & {\bf 3}  & {\bf 4} & {\bf 5}^a & {\bf 5}^b  \\[0.4cm] \hline
\frac{\pi}{\sqrt{3}} & \frac{\pi}{\sqrt{3}}\ln 3 & \frac{\pi}{\sqrt{3}}\ln^2 3 
&  \frac{\pi}{\sqrt{3}}\ln^3 3 & \frac{\pi}{\sqrt{3}}\ln^4  3
&  \frac{\pi}{\sqrt{3}} \zeta_4  \\[0.4cm] \hline
\ln3 & \zeta_2 & \zeta_2 \ln3  & \zeta_2 \ln^2 3  & \zeta_2 \ln^3 3 &  
\pi \zeta_2   \Cl{2}{\frac{\pi}{3}} \\[0.4cm] \hline
& \frac{\Cl{2}{\frac{\pi}{3}}}{\sqrt{3}} &
\frac{\Cl{2}{\frac{\pi}{3}}}{\sqrt{3}} \ln 3 & 
\frac{\Cl{2}{\frac{\pi}{3}}}{\sqrt{3}} \ln^2 3  & 
\frac{\Cl{2}{\frac{\pi}{3}}}{\sqrt{3}} \ln^3 3 &  
\frac{\pi}{\sqrt{3}} \left[ \Cl{2}{\frac{\pi}{3}} \right]^2 \\[0.4cm] \hline
& \ln^2 3 &  \frac{\pi}{\sqrt{3}} \zeta_2 & 
\frac{\pi}{\sqrt{3}} \zeta_2 \ln3 & 
\frac{\pi}{\sqrt{3}} \zeta_2  \ln^2 3  & 
\zeta_2 \frac{\Ls{3}{\frac{2 \pi}{3}}}{\sqrt{3}}  \\[0.4cm] \hline
&&  \pi  \Cl{2}{\frac{\pi}{3}} & \pi  \Cl{2}{\frac{\pi}{3}} \ln 3 & 
\pi  \Cl{2}{\frac{\pi}{3}}\ln^2 3   &  
\Cl{2}{\frac{\pi}{3}} \Ls{3}{\frac{2\pi}{3}}  \\[0.4cm] \hline
&&  \zeta_3                 & \zeta_3 \ln 3 & \zeta_3 \ln^2 3      & 
\pi \Ls{4}{\frac{ \pi}{3}}  \\[0.4cm] \hline
&&  \frac{\Ls{3}{\frac{2\pi}{3}}}{\sqrt{3}} & 
 \frac{\Ls{3}{\frac{2\pi}{3}}}{\sqrt{3}} \ln 3 & 
\frac{\Ls{3}{\frac{2\pi}{3}}}{\sqrt{3}} \ln^2 3 & 
\pi \Ls{4}{\frac{2 \pi}{3}}  \\[0.4cm] \hline
&& \ln^3 3 & \zeta_2 \frac{\Cl{2}{\frac{\pi}{3}}}{\sqrt{3}} & 
 \zeta_2 \frac{\Cl{2}{\frac{\pi}{3}}}{\sqrt{3}} \ln 3 &   
\zeta_3   \frac{\Cl{2}{\frac{\pi}{3}}}{\sqrt{3}} \\[0.4cm] \hline
&&& \frac{\pi}{\sqrt{3}} \zeta_3     & \frac{\pi}{\sqrt{3}} \zeta_3 \ln 3   &   
\zeta_2 \zeta_3    \\[0.4cm] \hline
&&& \pi \Ls{3}{\frac{2\pi}{3}} &  \pi \Ls{3}{\frac{2\pi}{3}}\ln 3   & 
\zeta_5    \\[0.4cm] \hline
&&&  \left[ \Cl{2}{\frac{\pi}{3}} \right]^2    &  
\left[\Cl{2}{\frac{\pi}{3}} \right]^2 \ln 3    &  
\frac{\pi}{\sqrt{3}} \LS{4}{1}{\frac{2\pi}{3}} \\[0.4cm] \hline
&&& \zeta_4                              &  \zeta_4 \ln 3    &    
\frac{\Cl{5}{\frac{\pi}{3}}}{\sqrt{3}} \\[0.4cm] \hline
&&& \frac{\Ls{4}{\frac{\pi}{3} }}{\sqrt{3}} &  
\frac{\Ls{4}{\frac{\pi}{3} }}{\sqrt{3}} \ln 3  & 
\frac{\Ls{5}{\frac{\pi}{3}}}{\sqrt{3}} \\[0.4cm] \hline
&&& \frac{\Ls{4}{\frac{2\pi}{3}}}{\sqrt{3}} & 
\frac{\Ls{4}{\frac{2\pi}{3}}}{\sqrt{3}} \ln 3  & 
\frac{\Ls{5}{\frac{2\pi}{3}}}{\sqrt{3}} \\[0.4cm] \hline
&&& \LS{4}{1}{\frac{2\pi}{3}} & 
\LS{4}{1}{\frac{2\pi}{3}} \ln 3  & 
\LS{5}{1}{\frac{2\pi}{3}} \\[0.4cm] \hline
&&& \ln^4 3 && \frac{\LS{5}{2}{\frac{2\pi}{3}}}{\sqrt{3}} \\[0.4cm] \hline
&&&&& \ln^5 3 \\[0.4cm] \hline
\end{array}
$$

\newpage

The above table contains the following functions, which appear in the
splitting of the polylogarithms into real and imaginary parts \cite{Lewin}.

\begin{eqnarray}
\Ls{n}{\theta} & = & -\int_0^\theta \ln^{n-1}
\left| 2 \sin \frac{\phi}{2} \right| d \phi ,
\nonumber \\
\LS{n}{m}{\theta} & = & -\int_0^\theta \phi^m \ln^{n-m-1}
\left| 2 \sin \frac{\phi}{2}\right| d \phi.
\nonumber 
\end{eqnarray}

\noindent
The basis of weight ${\bf 5}$ is sum of columns ${\bf 5}^a$
and ${\bf 5}^b$ in Table I. Some numerical values of these 
functions for special values of their argument are given in appendix A.

It should be noted, that the number $N_J$ of basis elements with weight J
satisfies the following relation:

\begin{equation}
N_J = 2 N_{J-1} \equiv 2^J.
\label{number}
\end{equation}

\section{Examples}

Let us present several examples of two- and three-loop diagrams
expressible in terms of the basis structure elaborated in the previous section.
Here we will use the notation
$S_2 \equiv \frac{4}{9\sqrt{3}}\Cl{2}{\frac{\pi}{3}}$ and are working in 
in Euclidean space-time with dimension $N= 4-2\varepsilon$. Moreover, 
the normalization factor 
$
\frac{\left( m^2 \right)^{\varepsilon} \left(4\pi \right)^{\frac{N}{2}}}{\Gamma(1+\varepsilon)}
$
for each loop is assumed.

As first example we consider the two-loop single mass scale bubble diagram 
({\bf VL111} in Fig.\ref{simple}).
Taking into account the `magic connection' \cite{magic} between the one-loop off-shell massless 
triangle and two-loop bubble integral and using the one-fold integral
representation  \cite{one} for the former, we have:

\begin{equation}
m^{-2} {\bf VL111}(1,1,1,m) =  -\frac{3}{2} 
\frac{1}{(1-\varepsilon)(1-2\varepsilon)}
\left \{ \frac{1}{\varepsilon^2} - \frac{1}{\varepsilon} 
\int\limits_0^1 \frac{dx  (1-x^{2\varepsilon}) }{(1-x+x^2)^{1+\varepsilon}}
\right \}.
\end{equation}

\noindent
Another form of the integral representation of this diagram can be found in 
\cite{two-loop}.
In this way, the calculation of the coefficients of the
$\varepsilon$-expansion is reduced to the following integrals:

\begin{equation}
I_{a,b} = \int\limits_0^1 \frac{d x}{1-x+x^2}
\ln^a x  \ln^b (1-x+x^2).
\label{I}
\end{equation}

\noindent
The values of these integrals for $a+b \leq 2$ are know from \cite{broadhurst1,master,one}.
The results for the next two indices ($a+b=3,4$)
are given in appendix B. Combining all these results we have:

\begin{eqnarray}
m^{-2} {\bf VL111}(1,1,1,m) 
=   -\frac{3}{2} \frac{1}{(1-\varepsilon)(1-2\varepsilon)}
\Biggl \{ \frac{1}{\varepsilon^2} - 9 S_2 
&& \nonumber \\
+ 2 \varepsilon \Biggl[
\frac{9}{2} S_2 \ln 3 - \frac{\pi}{\sqrt{3}} \zeta_2  
- 3 \frac{\Ls{3}{\frac{2 \pi}{3}}}{\sqrt{3}}
\Biggr]
&& \nonumber \\
+ 2 \varepsilon^2 \Biggl[
-\frac{9}{4} S_2 \ln^2 3 
+ \frac{\pi}{\sqrt{3}} \zeta_2 \ln 3
+ 3 \frac{\Ls{3}{\frac{2 \pi}{3}}}{\sqrt{3}} \ln 3
+ 2 \frac{\pi}{\sqrt{3}} \zeta_3
- 2 \frac{\Ls{4}{\frac{2 \pi}{3}}}{\sqrt{3}} 
\Biggr]
&& \nonumber \\
 + \varepsilon^3 \Biggl[
\frac{3}{2} S_2 \ln^3 3 
- \frac{\pi}{\sqrt{3}} \zeta_2 \ln^2 3
- 4 \frac{\pi}{\sqrt{3}} \zeta_3 \ln 3 
- \frac{19}{2} \frac{\pi}{\sqrt{3}} \zeta_4
- 3 \frac{\Ls{3}{\frac{2 \pi}{3}}}{\sqrt{3}} \ln^2 3
&& \nonumber \\
+ 
4 \frac{\Ls{4}{\frac{2 \pi}{3}}}{\sqrt{3}} \ln 3
- 2 \frac{\Ls{5}{\frac{2 \pi}{3}}}{\sqrt{3}}
\Biggr]
\Biggr \} +  {\cal O} (\varepsilon^4).
&& 
\end{eqnarray}

\noindent
This expression coincides with analytical results obtained before. The 
$\varepsilon$ -part was obtained in \cite{magic} and all higher terms
were obtained by Davydychev \cite{davydychev2}.

Let us consider another example: ${\bf J011}$ diagram of Fig.\ref{simple}
with the external momentum on the mass shell of the internal mass.
The first two terms of the $\varepsilon$-expansion of this
diagram has been calculated analytically in \cite{master2}. 
Here we present the next two terms of this expansion. 
It is well known that two diagrams with different sets of indices 
are needed as master integrals \cite{tarasov1}. For the numerical investigation we
choose the integral with indices 122 (the first index belonging to the
massless line) and a linear combination investigated in \cite{threshold}.
To obtain the numerical results for these two integrals with high precision
\cite{precision} we have constructed their Pad\'e approximants from the small
momentum Taylor expansion \cite{small}. 
The Taylor coefficients were obtained by means of the package
\cite{TLAMM}. Within the above mentioned basis we have:

\begin{eqnarray}
&&
m^2 {\bf J011}(1,2,2,m) =  \frac{2}{3} \zeta_2
- \varepsilon \frac{2}{3} \zeta_3 + \varepsilon^2 3 \zeta_4
- \varepsilon^3 \left \{ 2 \zeta_5 + \frac{4}{3} \zeta_2 \zeta_3
\right\}
\nonumber \\
&&
+ \varepsilon^4 \left \{ \frac{61}{6} \zeta_6 + \frac{2}{3} \zeta_3^2
\right\}
- \varepsilon^5 \left \{ 6 \zeta_7 + 4 \zeta_2 \zeta_5 + 6 \zeta_3
\zeta_4 \right\}
+ {\cal O} (\varepsilon^6),
\end{eqnarray}

\noindent
and

\begin{eqnarray}
&&
m^2 \Biggl[ {\bf J011}(1,2,2,m) + 2 {\bf J011}(2,1,2,m) \Biggr] =  
-\frac{2}{3 \varepsilon} \frac{\pi}{\sqrt{3}}
+ \Biggl\{ -9 S_2 + 2 \frac{\pi}{\sqrt{3}} \ln 3 \Biggr\}  
\nonumber \\
&& 
+ \varepsilon \Biggl\{
-3  \frac{\pi}{\sqrt{3}} \ln^2 3
+ 27 S_2 \ln 3
- \frac{28}{3} \frac{\pi}{\sqrt{3}} \zeta_2 
- 18 \frac{\Ls{3}{\frac{2\pi}{3}}}{\sqrt{3}} 
\Biggr\}
+ \varepsilon^2 \Biggl\{
- \frac{81}{8}S_2 \ln^2 3
\nonumber \\
&&
+ 3 \frac{\pi}{\sqrt{3}} \ln^3 3
+ 28 \frac{\pi}{\sqrt{3}} \zeta_2 \ln 3
+  \frac{112}{3} \frac{\pi}{\sqrt{3}} \zeta_3
+54 \frac{\Ls{3}{\frac{2\pi}{3}}}{\sqrt{3}} \ln 3
-36 \frac{\Ls{4}{\frac{2\pi}{3}}}{\sqrt{3}} 
+ \frac{32}{3} \frac{\Ls{4}{\frac{\pi}{3}}}{\sqrt{3}}
\Biggr\}
\nonumber \\ &&
+ \varepsilon^3 \Biggl\{
-\frac{9}{4} \frac{\pi}{\sqrt{3}} \ln^4 3
+ \frac{81}{2} S_2 \ln^3 3 
- 216 \zeta_2  \frac{\Ls{3}{\frac{2\pi}{3}}}{\sqrt{3}} 
- 81 \frac{\Ls{3}{\frac{2\pi}{3}}}{\sqrt{3}} \ln^2 3
- 32 \frac{\Ls{4}{\frac{\pi}{3}}}{\sqrt{3}} \ln 3
\nonumber \\ &&
+ 108 \frac{\Ls{4}{\frac{2 \pi}{3}}}{\sqrt{3}} \ln 3
+ \frac{196}{3} \frac{\Ls{5}{\frac{\pi}{3}}}{\sqrt{3}}
-54 \frac{\Ls{5}{\frac{2\pi}{3}}}{\sqrt{3}}
-42 \frac{\pi}{\sqrt{3}} \zeta_2 \ln^2 3
-112 \frac{\pi}{\sqrt{3}} \zeta_3 \ln 3
\nonumber \\ &&
- \frac{679}{6} \frac{\pi}{\sqrt{3}} \zeta_4
+ 108 \frac{\pi}{\sqrt{3}} \LS{4}{1}{\frac{2\pi}{3}}
- 81  \frac{\LS{5}{2}{\frac{2\pi}{3}}}{\sqrt{3}}
\Biggr\}
+ {\cal O} (\varepsilon^4), 
\end{eqnarray}

\noindent
The values of the master integrals (with indices 111 and 112) can be obtained
by means of recurrence relations \cite{massless1} implemented in 
the package {\bf ON-SHELL2}
\cite{ON-SHELL2}. They are relatively lengthy and therefore will not be presented
here.

\begin{figure}[t]
\centerline{\vbox{\epsfysize=30mm \epsfbox{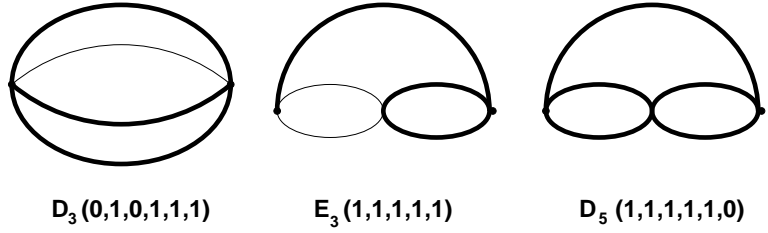}}}
\caption{\label{three} Three-loop master-integrals.
Bold and thin lines correspond to massive and
massless propagators, respectively.}
\end{figure}

Now we consider master integrals occurring in the package \cite{leo}:
$D_3(0,1,0,1,1,1,m)$, $E_3(1,1,1,1,1,m)$ and $D_5(1,1,1,1,1,0,m)$ as shown 
in Fig.\ref{three}, the notation being as in \cite{leo}. 
These integrals contribute in calculations like \cite{qcd}.

Results for $D_3(0,1,0,1,1,1,m)$ and $E_3(1,1,1,1,1,m)$ have not been
published before. 
Two independent approaches have been
applied to obtain high precision numerical values for these integrals.
One method started from analytical results \cite{broadhurst1,master2,FKV} for 
two-loop propagator diagrams. The finite three-loop diagrams were then obtained
by multiplication with the massive propagator with power $j>1$, resulting
in a one-fold integral representation, which was integrated by MAPLE
with accuracy more than 200 decimals.
Alternatively the  $E_3$-integral was calculated by the method which
was used earlier for the numerical calculation of $D_3(1,1,1,1,1,1)$ \cite{rho}.
Here the product of the two one-loop propagator diagrams were expanded 
in their small/large external momentum and the integration was performed piecewise.
Details of such calculation were published in \cite{expa}.
An accuracy of about 100 decimals was also obtained in this manner. 

The analytical results are the following ones:

\begin{eqnarray}
&& m^{-2} {\bf E_3}(1,1,1,1,1,m) = -\frac{2}{3 \varepsilon^3} - \frac{11}{3 \varepsilon^2}
+ \frac{1}{\varepsilon} \Biggl\{ - 14 + \frac{27}{2} S_2 - \zeta_2 \Biggr\}  
\nonumber \\ &&
+ \Biggl\{
- \frac{139}{3} 
+ \frac{135}{2} S_2 
- \frac{27}{2} S_2 \ln3
- 5 \zeta_2 - \frac{1}{3} \zeta_3
+ \frac{5}{3} \frac{\pi}{\sqrt{3}} \zeta_2
+ 9 \frac{\Ls{3}{\frac{2\pi}{3}}}{\sqrt{3}}
\Biggr\}  
\nonumber \\ &&
+ \varepsilon \Biggl\{
- \frac{430}{3}
+ \frac{459}{2} S_2
- \frac{135}{2} S_2 \ln 3
+ \frac{27}{4} S_2 \ln^2 3
+ 45 \frac{\Ls{3}{\frac{2\pi}{3}}}{\sqrt{3}}
- 9 \frac{\Ls{3}{\frac{2\pi}{3}}}{\sqrt{3}} \ln 3 
\nonumber \\ &&
+ \frac{80}{9} \frac{\Ls{4}{\frac{ \pi}{3}}}{\sqrt{3}}
+ 6 \frac{\Ls{4}{\frac{2\pi}{3}}}{\sqrt{3}}
- 17 \zeta_2 
+ \frac{25}{3} \frac{\pi}{\sqrt{3}} \zeta_2
- \frac{5}{3}  \frac{\pi}{\sqrt{3}} \zeta_2 \ln 3 
- \frac{13}{3} \zeta_3 
- \frac{94}{9} \frac{\pi}{\sqrt{3}} \zeta_3
\nonumber \\ &&
+ \frac{3}{2} \zeta_4
+ 9 S_2 \zeta_2
\Biggr\}
+ {\cal O} (\varepsilon^2),
\end{eqnarray}

\begin{eqnarray}
&& m^{-4} {\bf D_3}(0,1,0,1,1,1,m) = 
 \frac{1}{\varepsilon^3} + \frac{15}{4 \varepsilon^2}
+\frac{65}{8 \varepsilon}  + \frac{135}{16} 
+ \frac{81}{4} S_2
\nonumber \\ &&
+ \varepsilon \Biggl\{
- \frac{763}{32} 
+ \frac{1215}{8} S_2 
- \frac{243}{4} S_2 \ln 3 
+ \frac{27}{2} \zeta_2 \frac{\pi}{\sqrt{3}}
+ \frac{81}{2} \frac{\Ls{3}{\frac{2\pi}{3}}}{\sqrt{3}}
\Biggr\}
+ {\cal O} (\varepsilon^2).
\end{eqnarray}

Let us now consider the diagram $D_5(1,1,1,1,1,0)$ shown in
Fig.\ref{three}. As shown by Broadhurst \cite{master3} it
contains the new transcendental number $V_{3,1}$. We were
able to verify this, but we also found the following decomposition
in terms of our basis elements:

\begin{equation}
V_{3,1} = \frac{81}{16} S_2^2 - \frac{1}{4} \pi \Ls{3}{\frac{2\pi}{3}}+
\frac{13}{24} \zeta_3 \ln 3
- \frac{259}{108} \zeta_4 + \frac{3}{8} \LS{4}{1}{\frac{2\pi}{3}}.
\label{v31}
\end{equation}

\section{Conclusion}

In this work we have constructed the basis of transcendental numbers
for single mass scale diagrams with two-particle massive cuts.

The basis , up to weight 5, is given in Table I. The conjectures which
have been used for the construction of the basis are as follows: the set 
$\{b_J \}$ of transcendental numbers of weight J has the structure: 

$$
b_J = \{b_{J-K} b_K, \tilde b_J \}, K=1,2, \cdots, J-1,
$$

\noindent 
where new elements $\{ \tilde b_J\} $ are connected with functions
arising in the splitting of polylogarithms  $\Li{J}{z}$, $\Li{J}{1-z}$ 
into real and imaginary parts, and
$z=exp\left( i \frac{\pi}{3} \right), exp\left( i \frac{2 \pi}{3} \right)$ . 
Up to the weight {\bf 5} we observed also, that the 
the number $N_J$ of basis elements with weight J
satisfies the relation (\ref{number}). 

The fact that the `sixth root of unity' \cite{master3} plays such a crucial
role in the construction of our basis remains mysterious.
The introduction of the `normalization factor'
$\frac{1}{\sqrt{3}}$ remains a matter of experience.
Nevertheless our construction allows us to calculate
several new diagrams (or obtain new terms in
the $\varepsilon$-expansion) in an analytical form.

The interesting question remaining beyond our work is how to construct
the basis for single-mass scale diagrams having several different
massive cuts. From multiloop QED and QCD calculations 
\cite{broadhurst2,master3,massless2,FJTV,amm} we know that other transcendental
numbers, like $\pi,\ln2, G,\Li{n}{\frac{1}{2}}$ or their products appear.
In this case the structure of the basis elements is connected with polylogarithms of 
$z=exp\left( i \frac{\pi}{2} \right), exp\left( i \pi \right)$ \cite{master3}.  
At the present moment we don't know how the mixing basis could look like.

\noindent
{\bf Acknowledgments}
We are grateful to A.~Davydychev and O.~V.~Tarasov for useful 
discussions and careful reading of the manuscript and to  
M.~Tentyukov and  O.~Veretin for help in numerical calculation.
We are very indebted to A.~Davydychev for informing us about the 
results of \cite{davydychev2} before publication. 
M.K's research has been supported by the DFG project FL241/4-1 
and in part by RFBR $\#$98-02-16923.

\setcounter{section}{0}
\setcounter{equation}{0}
\renewcommand{\theequation}{A.\arabic{equation}}
\renewcommand{\thesection}{APPENDIX A. }
\section{Numerical values of basic elements.}

Here we present numerical values up to 40 decimals of functions, 
which for special
values of their argument provide our basic elements. These numerical
values were obtained by numerical integration of their integral 
representation by means of MAPLE.

\begin{eqnarray}
\Cl{2}{\frac{\pi}{3}} &=&
  1.0149416064096536250212025542745202859417\cdots,
\nonumber \\
\Ls{3}{\frac{2\pi}{3}} &=& 
 -2.1447672125694943914826421122153512974711\cdots,
\nonumber \\
\Ls{4}{\frac{\pi}{3}}  &=& 
  6.0094975498188889162047887062032707405970\cdots,
\nonumber \\
\Ls{4}{\frac{2 \pi}{3}} &=& 
  5.9506682441397849307578101290139840215916\cdots,
\nonumber \\
\LS{4}{1}{\frac{2 \pi}{3}} &=&
 -0.4976755516066471960582989542950996307358\cdots,
\nonumber \\
\Cl{5}{\frac{\pi}{3}} &=& 
  0.4800591458997082992274840215078861889153\cdots,
\nonumber \\
\Ls{5}{\frac{\pi}{3}} &=& 
-24.0125331255169146150157139636316267950288\cdots,
\nonumber \\
\Ls{5}{\frac{2 \pi}{3}} &=& 
-24.0394144085686386066499683097012073111734\cdots,
\nonumber \\
\LS{5}{1}{\frac{2 \pi}{3}} &=& 
  0.2708016246010793737741869955751992363356\cdots,
\nonumber \\
\LS{5}{2}{\frac{2 \pi}{3}} &=& 
 -0.5181087868296801173472656387316967550219\cdots.
\nonumber 
\end{eqnarray}
\setcounter{section}{0}
\setcounter{equation}{0}
\renewcommand{\theequation}{B.\arabic{equation}}
\renewcommand{\thesection}{APPENDIX B. }
\section{Analytical results of integrals occurring in the
$\varepsilon$-expansion of the two-loop bubble.}

Here we present the values of the integrals (\ref{I}). Their numerical
values were obtained by MAPLE with accuracy of 230 decimal and again
the analytical results were obtained by PSLQ.

\begin{eqnarray}
I_{3,0} & = & -\frac{8}{3} \frac{\Ls{4}{\frac{\pi}{3}}}{\sqrt{3}} 
+ \frac{4}{3} \frac{\pi}{\sqrt{3}} \zeta_3,
\nonumber \\
I_{2,1} & = & \zeta_2 S_2 -\frac{248}{81} \frac{\Ls{4}{\frac{\pi}{3}}}{\sqrt{3}} 
+ \frac{20}{27} \frac{\pi}{\sqrt{3}} \zeta_2 \ln 3 
+ \frac{280}{81} \frac{\pi}{\sqrt{3}} \zeta_3,
\nonumber \\
I_{1,2} & = & 2 \zeta_2 S_2 -\frac{9}{2} S_2 \ln^2 3 
+ 6\frac{\Ls{3}{\frac{2\pi}{3}}}{\sqrt{3}} \ln3 
-\frac{208}{81} \frac{\Ls{4}{\frac{\pi}{3}}}{\sqrt{3}} 
- 4 \frac{\Ls{4}{\frac{2\pi}{3}}}{\sqrt{3}} 
\nonumber \\
&& 
+ \frac{94}{27} \frac{\pi}{\sqrt{3}} \zeta_2 \ln 3 
+ \frac{740}{81} \frac{\pi}{\sqrt{3}} \zeta_3,
\nonumber \\
I_{0,3} & = & \frac{2}{3} \frac{\pi}{\sqrt{3}} \ln^3 3 
- 18 S_2 \ln^2 3 
+ 24 \frac{\Ls{3}{\frac{2\pi}{3}}}{\sqrt{3}} \ln3 
- 16 \frac{\Ls{4}{\frac{2\pi}{3}}}{\sqrt{3}} 
\nonumber \\
&& 
+ 12 \frac{\pi}{\sqrt{3}} \zeta_2 \ln 3 
+ 24  \frac{\pi}{\sqrt{3}} \zeta_3,
\label{I3}
\end{eqnarray}

\noindent
and 

\begin{eqnarray}
I_{4,0} & = & \frac{340}{27} \frac{\pi}{\sqrt{3}} \zeta_4,
\nonumber \\
I_{3,1} & = & - 6 S_2 \zeta_3 - 18 \zeta_2 \frac{\Ls{3}{\frac{2\pi}{3}}}{\sqrt{3}}
-\frac{8}{3} \frac{\Ls{4}{\frac{\pi}{3}}}{\sqrt{3}} \ln 3
+ \frac{49}{9} \frac{\Ls{5}{\frac{\pi}{3}}}{\sqrt{3}} 
\nonumber \\ && 
+ \frac{4}{3} \frac{\pi}{\sqrt{3}} \zeta_3 \ln 3
+ \frac{5957}{216} \frac{\pi}{\sqrt{3}} \zeta_4
+ 9 \frac{\pi}{\sqrt{3}} \LS{4}{1}{\frac{2\pi}{3}}
- \frac{27}{4} \frac{\LS{5}{2}{\frac{2 \pi}{3}}}{\sqrt{3}},
\nonumber \\
I_{2,2} & = & 2 S_2 \zeta_2 \ln 3 - \frac{46}{3} S_2 \zeta_3 
+ 9 \frac{\pi}{\sqrt{3}} S_2^2 
- \frac{128}{3} \zeta_2 \frac{\Ls{3}{\frac{2\pi}{3}}}{\sqrt{3}}
- \frac{496}{81} \frac{\Ls{4}{\frac{\pi}{3}}}{\sqrt{3}} \ln 3
\nonumber \\ && 
+ \frac{986}{81} \frac{\Ls{5}{\frac{\pi}{3}}}{\sqrt{3}} 
+ \frac{20}{27} \frac{\pi}{\sqrt{3}} \zeta_2 \ln^2 3
+ \frac{560}{81} \frac{\pi}{\sqrt{3}} \zeta_3 \ln 3
+ \frac{16963}{324} \frac{\pi}{\sqrt{3}} \zeta_4
\nonumber \\ && 
+ \frac{62}{3} \frac{\pi}{\sqrt{3}} \LS{4}{1}{\frac{2\pi}{3}}
- \frac{31}{2} \frac{\LS{5}{2}{\frac{2 \pi}{3}}}{\sqrt{3}},
\nonumber \\
I_{1,3} & = & 6 S_2 \zeta_2 \ln 3 - 22 S_2 \zeta_3 
+ 27 \frac{\pi}{\sqrt{3}} S_2^2 - \frac{9}{2} S_2 \ln^3 3
- 56 \zeta_2 \frac{\Ls{3}{\frac{2\pi}{3}}}{\sqrt{3}}
+ 9 \frac{\Ls{3}{\frac{2 \pi}{3}}}{\sqrt{3}} \ln^2 3
\nonumber \\ && 
- \frac{208}{27} \frac{\Ls{4}{\frac{\pi}{3}}}{\sqrt{3}} \ln 3
- 12 \frac{\Ls{4}{\frac{2 \pi}{3}}}{\sqrt{3}} \ln 3
+ \frac{398}{27} \frac{\Ls{5}{\frac{\pi}{3}}}{\sqrt{3}} 
+ 6 \frac{\Ls{5}{\frac{2 \pi}{3}}}{\sqrt{3}} 
\nonumber \\ && 
+ \frac{47}{9} \frac{\pi}{\sqrt{3}} \zeta_2 \ln^2 3
+ \frac{740}{27} \frac{\pi}{\sqrt{3}} \zeta_3 \ln 3
+ \frac{10847}{108} \frac{\pi}{\sqrt{3}} \zeta_4
\nonumber \\ && 
+ 26 \frac{\pi}{\sqrt{3}} \LS{4}{1}{\frac{2\pi}{3}}
- \frac{39}{2} \frac{\LS{5}{2}{\frac{2 \pi}{3}}}{\sqrt{3}},
\nonumber \\
I_{0,4} & = & \frac{2}{3} \frac{\pi}{\sqrt{3}} \ln^4 3 
- 24 S_2 \ln^3 3 
+ 48 \frac{\Ls{3}{\frac{2\pi}{3}}}{\sqrt{3}} \ln^2 3 
- 64 \frac{\Ls{4}{\frac{2\pi}{3}}}{\sqrt{3}} \ln 3 
\nonumber \\
&& 
+ 32 \frac{\Ls{5}{\frac{\pi}{3}}}{\sqrt{3}} 
+ 24 \frac{\pi}{\sqrt{3}} \zeta_2 \ln^2 3 
+ 96  \frac{\pi}{\sqrt{3}} \zeta_3 \ln3 
+ 228 \frac{\pi}{\sqrt{3}} \zeta_4.
\label{I4}
\end{eqnarray}

\end{document}